\documentstyle[12pt,twoside,epsf]{article}
\begin{document}

\newcommand{\Lslash}[1]{ \parbox[b]{1em}{$#1$} \hspace{-0.8em}
                         \parbox[b]{0.8em}{ \raisebox{0.2ex}{$/$} }    }
\newcommand{\Sslash}[1]{ \parbox[b]{0.6em}{$#1$} \hspace{-0.55em}
                         \parbox[b]{0.55em}{ \raisebox{-0.2ex}{$/$} }    }
\newcommand{\Mbf}[1]{ \parbox[b]{1em}{\boldmath $#1$} }
\newcommand{\mbf}[1]{ \parbox[b]{0.6em}{\boldmath $#1$} }
\newcommand{\beq}{\begin{equation}}
\newcommand{\eeq}{\end{equation}}
\newcommand{\beqa}{\begin{eqnarray}}
\newcommand{\eeqa}{\end{eqnarray}}
\newcommand{\skipfields}{\!\!\!\!\! & \!\!\!\!\! &}

\newcommand{\gsim}{\buildrel > \over {_\sim}}
\newcommand{\lsim}{\buildrel < \over {_\sim}}
\newcommand{\ie}{{\it i.e.}}
\newcommand{\eg}{{\it e.g.}}
\newcommand{\cf}{{\it cf.}}
\newcommand{\etal}{{\it et al.}}
\newcommand{\gev}{{\rm GeV}}
\newcommand{\jpsi}{J/\psi}
\newcommand{\order}[1]{${\cal O}(#1)$}
\newcommand{\eq}[1]{eq.\ (\ref{#1})}

\newcommand{\ptr}{p_T}
\newcommand{\as}{\alpha_s}

\newcommand{\RS}{R_S(0)}
\newcommand{\Pqg}{P_{{\!}_{q\to g}}}
\newcommand{\psq}{\vec{p}_\perp^{\; 2}}

%
%***** For Document style: article
\newcommand{\LevelOne}[1]{ \subsection*{#1} }
\newcommand{\LevelTwo}[1]{ \subsubsection*{#1} }
\newcommand{\LevelThree}[1]{ \paragraph*{#1} }
\newcommand{\AppendixLevel}[2]{
%   \newpage
   \subsection*{Appendix #1 -- #2}
   \markright{Appendix #1}
   \addcontentsline{toc}{subsection}{\protect\numberline{#1}{#2}} }

\begin{titlepage}
\begin{flushright}
        NORDITA-96/3 P
\end{flushright}
\begin{flushright}
        hep-ph/9601333 \\ January 25, 1996  %% \today
\end{flushright}
\vskip .8cm
\begin{center}
{\Large Charmonium Production via Fragmentation

at Higher Orders in $\as$}
\vskip .8cm
{\bf P. Ernstr\"om, P. Hoyer, and M. V\"anttinen}
\vskip .5cm
NORDITA\\
Blegdamsvej 17, DK-2100 Copenhagen \O
\vskip 1.8cm
\end{center}
\begin{abstract}
\noindent Quarkonium production at a given large $\ptr$ is dominated by
parton fragmentation: a parton which is produced
with transverse momentum $\ptr/z$
fragments into a quarkonium state which carries a fraction $z$
of the parton momentum. Since parton production cross sections fall
steeply with $\ptr$, high $z$ fragmentation is favored. However,
quantum number constraints may require the emission of gluons in
the fragmentation process, and this softens the $z$ dependence of
the fragmentation function. We discuss the possibility that higher-order
processes may enhance the large $z$ part of fragmentation functions
and thus contribute significantly to the quarkonium cross section.
An explicit calculation of light quark fragmentation into $\eta_c$
shows that the higher-order process $q \rightarrow q\eta_c$ in fact
dominates the lowest-order process $q \rightarrow qg\eta_c$.
\end{abstract}
\end{titlepage}

\setlength{\baselineskip}{7mm}
\raggedbottom

\subsection*{1. Introduction}

Heavy quark production is a hard QCD process, and the perturbative
expansion of production amplitudes in $\as$ is expected to apply.
However, the perturbation series is not always
dominated by its lowest-order term.
For example, when the transverse momentum $\ptr$ of a collinear heavy
quark pair is much larger than its invariant mass $M$, light
parton fragmentation dominates the lowest-order production processes by
a power of $\ptr^2/M^2$ \cite{BraatenYuan}.
The data on charmonium production indeed shows a
$1/\ptr^4$ behavior of the cross section \cite{CDF,D0},
compatible with the prediction for a fragmentation process.

QCD calculations based on the color singlet model \cite{Baier}
nevertheless disagree
with data on the relative production rates of $S$ and $P$ wave
quarkonia and on the absolute normalization of the cross sections
at both low \cite{Schuler,VHBT} and high \cite{CDF,D0,highptTheory} values of
$\ptr$.
Order-of-magnitude discrepancies have been observed for
both charmonium and bottomonium.

In QCD, the fragmentation of a virtual gluon into a $^3S_1$
quarkonium state ($\jpsi,\ \Upsilon$) requires the emission of at least two
extra gluons, $g^*\rightarrow \mbox{}^3S_1 +gg$, whereas fragmentation into a
$^3P_J$ state can proceed with the emission of a single gluon,
$g^* \rightarrow \mbox{}^3P_J +g$.
The emission of the extra gluon suppresses the calculated $^3S_1$
cross section considerably compared to the $^3P_J$ cross section,
due to the extra power of $\as$ and also because the emitted
gluons carry away part of the transverse momentum of the fragmenting gluon.
The experimental $^3S_1$/$^3P_2$ cross section ratio
is much larger than the calculated one.

The discrepancy could be related to the bound state dynamics of
the quarkonia.
This is the solution proposed by the color octet model
\cite{BraatenFleming,BBL}. The gluon is assumed to first fragment
into a $Q \bar Q$ pair in a color octet state.
Later, after a formation time
characteristic of the bound state, the pair couples to the
physical quarkonium through the absorption or emission of one
or two soft gluons.
The momenta of the emitted gluons are typical
of the quarkonium bound state dynamics, and thus they
carry away only a minor part of the transverse momentum.

Although the probabilities of these nonperturbative transitions
are essentially free parameters, the octet model makes specific predictions
about the polarization of the produced quarkonium
\cite{Wise}.
The $^3S_1$ quarkonia produced by the fragmentation of a nearly real gluon 
are expected to be transversely polarized.
The data to test this prediction is not yet available.
Note that theoretical calculations of quarkonium polarization can be
compared with data in fixed-target experiments.
In this case, neither the color singlet model nor the color octet model
can explain the experimentally observed unpolarized production \cite{VHBT,TV}.

Here we wish to draw attention to the possibility that
higher-order perturbative contributions to fragmentation
mechanisms could be enhanced by the so-called trigger bias effect.
When the transverse momentum $\ptr$ of the quarkonium is fixed, processes
which allow the fragmenting parton to be produced with the lowest possible
transverse momentum $\ptr/z$ are favored.
The large $\ptr$ cross section is a convolution of the production cross
section $\sigma_i$ of a parton $i$ and its fragmentation function
$D_{i\rightarrow {\cal O}}$ to the quarkonium state ${\cal O}$,
\beq
  {d\sigma_{\cal O}(s,p_T) \over dp_T}
  = \sum_i \int_0^1 dz
    {d\sigma_i \over d{p_i}_T}(s,p_T/z,\mu)
    D_{i \rightarrow {\cal O}}(z,\mu) ,
\label{sigma}
\eeq
where $\mu$ is the factorization scale.
The parton production cross section $d\sigma_i / d{p_i}_T$
falls approximately as $(p_T/z)^{-4}$, which implies the enhancement of
high $z$ fragmentation by a factor $z^4$.
As a rough measure of the importance of a
fragmentation function ${D}_{i \rightarrow {\cal O}}(z,\mu)$
we can therefore use its
fifth moment ${D}^{(5)}_{i \rightarrow {\cal O}}(\mu)$, defined by
\beq
{D}^{(n)}_{i \rightarrow {\cal O}}(\mu) \equiv
\int_0^1 dz z^{n-1} D_{i \rightarrow {\cal O}}(z,\mu). \label{momentdef}
\eeq

\begin{figure}[htbp]
\begin{center}
\leavevmode
{\epsfxsize=13truecm \epsfbox{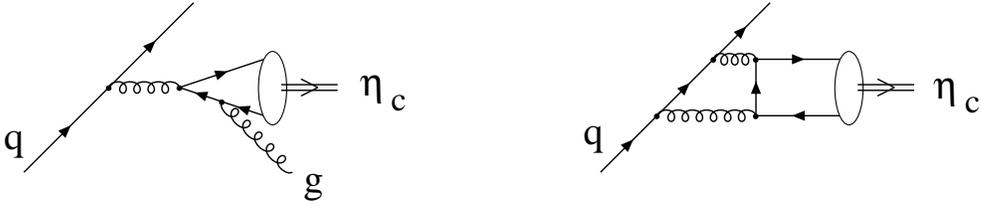}}
\end{center}
\caption[*]{
(a) A lowest-order diagram contributing to the process
$q \to \eta_c + X$. (b) A higher-order diagram with no gluon emission.
In each case there is another diagram with the $c$-quark--gluon
vertices interchanged.}
\label{qtoeta}
\end{figure}

We have studied the importance of higher-order fragmentation contributions in
the case of light quark fragmention into $^1S_0$ quarkonium ($\eta_c$)
within the color singlet model.
The relevant Feynman diagrams are shown in Fig.\ \ref{qtoeta}.
In the nonrelativistic limit the fragmentation function is a product of
the fragmentation probability into a collinear, on-shell $c\bar c$ pair in a
$^1S_0$ state and the square of the $\eta_c$ wave function at $r=0$,
\beq
  D_{i\rightarrow \eta_c}(z,\mu) =
  D_{i\rightarrow c\bar{c}(^1S_0)}(z,\mu) { \left|\RS\right|^2 \over 4\pi }.
  \label{SingletModel}
\eeq

\vspace{5mm}

At lowest order, $q \rightarrow \eta_c +X$ fragmentation is due to the
process $q \rightarrow q\eta_cg$ of Fig.\ \ref{qtoeta}a.
The emission of the gluon
suggests that this process may have a softer fragmentation function
than the higher-order process $q \rightarrow q\eta_c$
shown in Fig.\ \ref{qtoeta}b.
Due to the trigger bias effect, the
higher-order process could be enhanced.

%\pagebreak
\vspace{1cm}

\subsection*{2. Results}

Our calculation of the $q \to \eta_c +X$ fragmentation processes
shown in Fig.\ \ref{qtoeta} is described in the Appendix.
The contribution of the lowest-order process (Fig.\ \ref{qtoeta}a)
to the fragmentation function is of the form
\beq
  D^{(a)}_{{\!\!}_{q\to \eta_c}}(z,\mu) =
  f(z) \ln \left( {\mu^2 \over 4m_c^2 } \right) + g(z) +
  {\cal O} \left({4m_c^2 \over \mu^2} \right) .
  \label{LOres}
\eeq
The coefficient functions are
\beqa
  f(z) & = & {\alpha_s^3 C_F \left| \RS \right|^2 \over 48 \pi^2 m_c^3 N_c}
             \Biggl\{
             6\,\left( 2 + z \right)
             \left[ {{\pi }^2\over 6}-{\cal L}_2(z) \right]
             - 3\,z\,\ln (z)
             \nonumber \\
       &   & \mbox{} 
             + {2\over z} - 18 + 12\,z + 4\,{z^2}
             + \left( 6 + {6\over z} - 12\,z \right) \,\ln (1 - z)
             \Biggr\} ,
             \label{fexpr}
             \\
  g(z) & = & {\alpha_s^3 C_F \left| \RS \right|^2 \over 48 \pi^2 m_c^3 N_c}
             \Biggl\{
             {{\pi }^2}\,\left( 2 + z \right) \,\ln (z) 
             + \left( -18 + {2\over z} + 18\,z + 4\,{z^2} \right) \,\ln (z)
             \nonumber \\
       &   & \mbox{} +
             \left( 6 + {6\over z} - 12\,z \right) \,\ln (1 - z)\,\ln (z)
             - 3\,z\,{{\ln (z)}^2}
             + 34 + {{\pi }^2}\,\left( {1\over z} - 2 \right)
             \nonumber \\
       &   & \mbox{} - {{71}\over {6\,z}}
             - {{53\,z}\over 2} + {{13\,{z^2}}\over 3}
             + \left( {{-5}\over z} + 9\,z - 4\,{z^2} \right) \,\ln (1 - z)
             \nonumber \\
       &   & \mbox{} + \left( 18 - 12\,z \right) \,{\cal L}_2(z)
             + 6\,\left( 2 + z \right) \, \left[
             {\cal L}_3(z) - \ln (z)\,{\cal L}_2(z) - \zeta (3) \right]
             \Biggr\} ,
             \label{gexpr}
\eeqa
where
\beq
  {\cal L}_2(z) = - \int_0^z \frac{\ln(1-t)}{t} \; dt
\eeq
is the dilogarithmic function,
\beq
  {\cal L}_3(z) = \int_0^z \frac{{\cal L}_2(t)}{t} \; dt,
\eeq
and $\zeta(3) \approx 1.202$. The logarithmic term
$f(z) \ln [\mu^2/ (4m_c^2) ]$ arises from
the two-step process where $q \to qg$ splitting is
followed by $g \to \eta_cg$ fragmentation; the function
$f(z)$ can be written as 
\beq
  f(z) = \int_z^1 {dy \over y}
         \Pqg(z/y) D_{{\!\!}_{g\to \eta_c}}(y) ,
         \label{gconv}
\eeq
where $\Pqg(z/y)$ is the standard $q \to qg$ splitting
function \cite{AP} and $D_{{\!\!}_{g\to \eta_c}}(y)$ is the $g \to \eta_cg$
fragmentation function at lowest order \cite{BraatenYuan}.
A similar result has been obtained in
the case of $J/\psi$ production by light quark
fragmentation \cite{Cheung,Cho}.

A lower limit of the loop contribution (see Fig.\ \ref{qtoeta}b)
is obtained by considering only the imaginary part of the loop amplitude.
There is no logarithmic term in this case:
\beq
  D^{(b)}_{{\!\!}_{q\to \eta_c}}(z,\mu) \ge
  h(z) + {\cal O}\Bigl({4m_c^2 \over \mu^2}\Bigr),
  \label{HOres}
\eeq
where
\beqa
  h(z) & = & {\alpha_s^4 \left| \RS \right|^2 C_F^2 \over 96 \pi m_c^3 N_c}
             \Biggl\{
             14(1-z) \left[ {\pi^2 \over 6} - {\cal L}_2(1-z) \right]
             \nonumber \\
       &   & \mbox{} + z + {2z \over 1-z}\ln(z)
             + {z(7z^2-18z+12) \over (1-z)^2}\ln^2(z)
             \Biggr\} .
             \label{hexpr}
\eeqa

\begin{figure}[htbp]
\begin{center}
\leavevmode
{\epsfxsize=13truecm \epsfbox{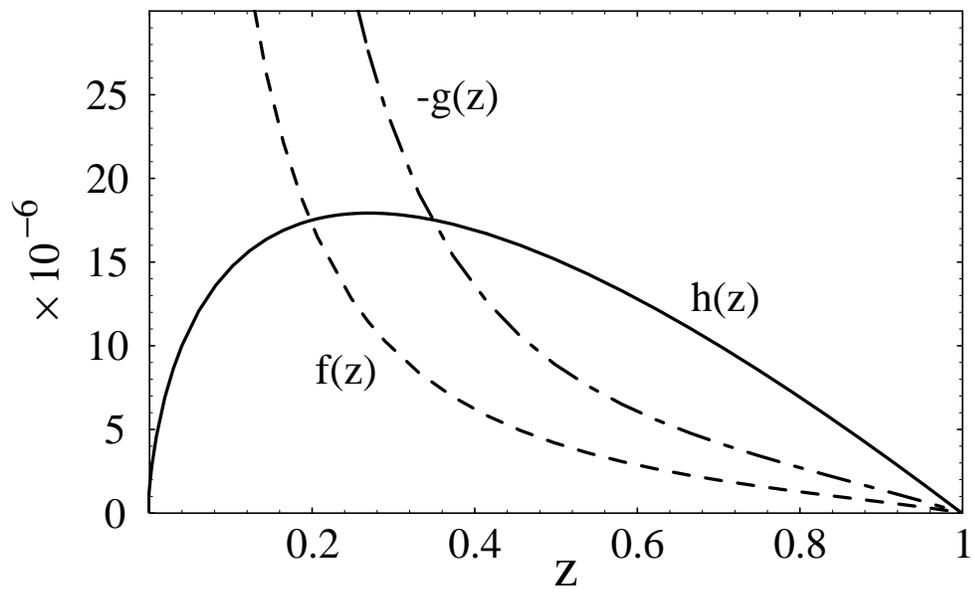}}
\end{center}
\caption[*]{
The functions $f(z)$, $-g(z)$ and $h(z)$ as defined in the text.}
\label{fgh}
\end{figure}

The functions $f(z)$, $-g(z)$ and $h(z)$ are plotted in Fig.\ \ref{fgh},
using \mbox{$\as=0.26$}, $|\RS|^2=(0.8 \;\gev)^3$, and $m_c=1.5 \;\gev$.
The loop contribution dominates
over the lower-order Born contribution for
$z \gsim 0.3$, even though the real part of the loop was neglected.
More quantitatively, the fifth moments of the Born and loop contributions
have the numerical values
\beqa
  \skipfields
  {D}^{(5,a)}_{q\to \eta_c}(\mu)
    = \int_0^1 dz z^4 D_{q\to \eta_c}(z,\mu)
    \approx \left[ 2.4 \ln \left( {\mu^2 \over 4m_c^2 } \right) - 5.1 \right]
            \times 10^{-7} ,
  \\ \skipfields
  {D}^{(5,b)}_{q\to \eta_c}(\mu)
    \ge \int_0^1 dz z^4 h(z) \approx 1.1 \times 10^{-6}.
  \label{fifthmom}
\eeqa
Depending on the fragmentation scale $\mu$, the contribution
from the loop diagram is thus up to an order of magnitude larger
than the lowest-order Born contribution.
Neglecting the higher-order process would lead to a major underestimate
of the fragmentation cross section.

It is possible to further simplify the calculation of the fragmentation
functions by taking advantage of the fact that only the large $z$ region is
important, due to the trigger bias effect.
We have verified that using only the leading part of an expansion of
$D(z,\mu)$ around $z=1$ changes the fifth moments of the loop and Born
contributions by less than 10\%.

%\pagebreak
\vspace{1cm}

\subsection*{3. Discussion}

The trigger bias effect in large $\ptr$ quarkonium production favors
fragmentation processes where the quarkonium takes a large fraction
$z$ of the momentum of the fragmenting parton. 
When estimating the relative importance of
different fragmentation processes, the shape of their $z$ dependence
must therefore be considered.

In particular, some higher-order perturbative contributions
may be enhanced relative to the lowest-order contributions
due to the trigger bias effect.
In this paper, we analyzed the process $q \to \eta_c + X$, where such
an enhancement can be expected because gluon emission is not required
in higher-order processes.
We found that there is a loop contribution which
indeed dominates the Born contribution by a large factor.

\begin{figure}[htbp]
\begin{center}
\leavevmode
{\epsfxsize=13truecm \epsfbox{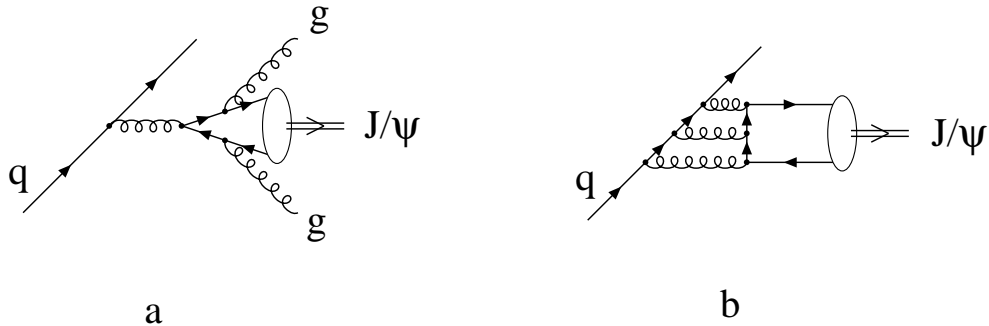}}
\end{center}
\caption[*]{
Light quark fragmentation into a $\jpsi$.}
\label{qtoJpsi}
\end{figure}

It is likely that an analogous result is obtained in the case
of $q \to \jpsi$ fragmentation.
Some relevant Born and loop diagrams are shown in Fig.\ \ref{qtoJpsi}.
At higher orders, all the gluons coupling to the heavy quark line can
be attached to the light quark line instead of being emitted, which
suggests a hard $z$ dependence of the fragmentation function.

These higher-order contributions are part of the standard perturbation
series and thus do not bring in any new parameters. Their relative
importance should depend only weakly on the quark mass (through the
decrease of $\as(m_Q)$ with $m_Q$). This is in qualitative agreement
with total cross section data \cite{CDF,D0,Schuler,VHBT}, which shows
a disagreement with Born term calculations (within the colour singlet
model) of similar magnitude for bottomonium and for charmonium.

\begin{figure}[htbp]
\begin{center}
\leavevmode
{\epsfxsize=13truecm \epsfbox{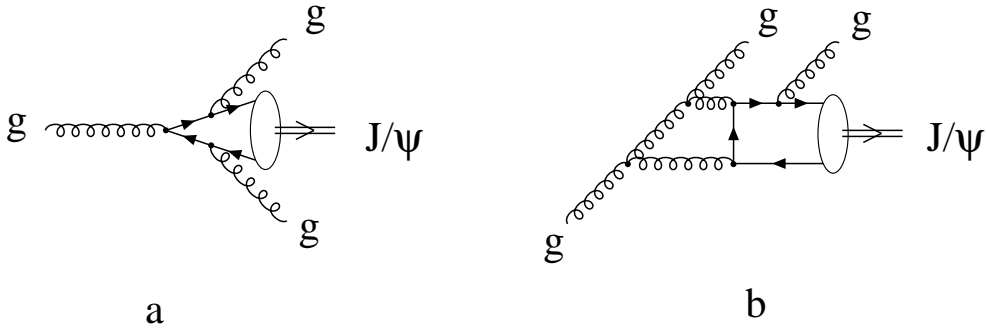}}
\end{center}
\caption[*]{
Gluon fragmentation into a $\jpsi$. (a) A lowest-order diagram. (b) A higher
order diagram.}
\label{gtoJpsi}
\end{figure}

The calculation presented here is not, however, immediately applicable
to the present data on quarkonium production. The primary production
mechanism for quarkonia at large $\ptr$
in hadron collisions is expected to be gluon
fragmentation. Even at higher orders, a minimum of two extra gluons
need to accompany a produced $\jpsi$, due to charge conjugation invariance
(\cf\ Fig.\ \ref{gtoJpsi}). In this case, loop diagrams like the one in
Fig.\ \ref{gtoJpsi}b simply represent radiative corrections to the
lowest-order process. Whether they enhance the kinematic region
where the emitted gluons carry little momentum (the large $z$ region)
can only be determined by an explicit calculation.

On the other hand, processes such as the one in Fig.\ \ref{qtoJpsi}b
could be significant in collisions where light quarks are more
copiously produced relative to gluons, such as at HERA. There,
however, also charm quark fragmentation becomes important as a
charmonium production mechanism at large $\ptr$ \cite{Godbole}.

In summary, we have pointed out that the trigger bias enhancement
of large $z$ fragmentation is crucial in quarkonium production
at large $\ptr$. As a specific example, we considered the
$q \to \eta_c$ fragmentation process and calculated a higher-order
perturbative correction whose contribution to the cross section
exceeds the lowest-order fragmentation contribution by a large
factor.

\bigskip

{\bf Acknowledgement.} We are grateful for
discussions with Stan Brodsky.

%\pagebreak
\vspace{1cm}

\subsection*{Appendix}

\begin{figure}[htbp]
\begin{center}
\leavevmode
{\epsfxsize=8truecm \epsfbox{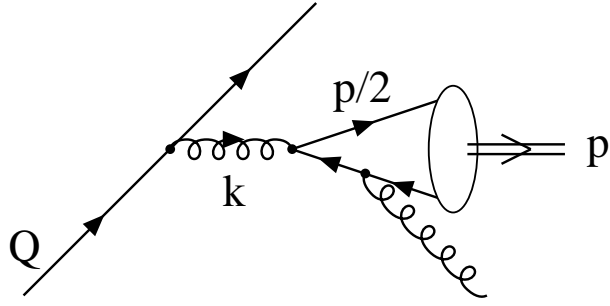}}
\end{center}
\caption[*]{
Momentum definitions in light quark fragmentation into $\eta_c$.}
\label{momentumdef}
\end{figure}

\begin{figure}[htbp]
\begin{center}
\leavevmode
{\epsfxsize=13truecm \epsfbox{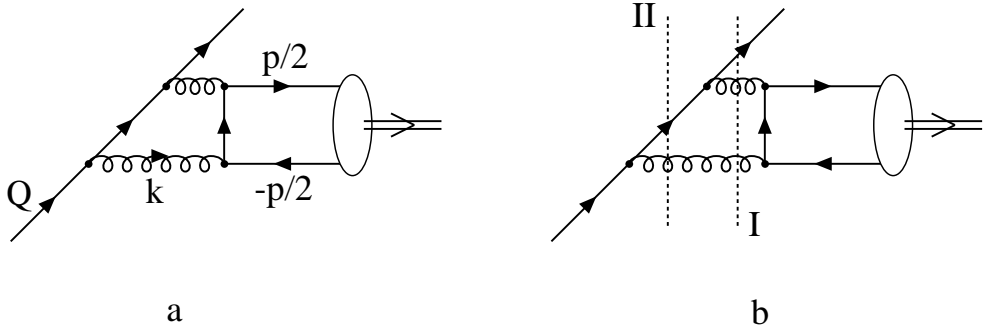}}
\end{center}
\caption[*]{
(a) Momentum definitions for the loop diagram contribution to $\eta_c$
production. (b) The cuts which give the imaginary part
of the loop diagram.}
\label{loopdef}
\end{figure}

We describe here our calculation of
the $q \to \eta_c$ fragmentation functions. As shown in
Figs.\ \ref{momentumdef} and \ref{loopdef}a, we denote by 
$p$ the momentum of the quarkonium state;
$Q$ denotes the momentum of the fragmenting light quark,
and $s=Q^2$ is its virtuality.

We work in the center of mass system of the light quark production process,
and choose the third axis along the direction of the fragmenting light quark.
The light-cone components of a four-vector $v$ are defined as
$v^\pm = v^0 \pm v^3$ and $\vec{v}_\bot=(v^1,v^2)$.
The variable $z$ is defined as $z=p^+/Q^+$, which in the limit of
large $Q^+$ is the fraction of the light quark momentum taken by
the $\eta_c$.

We use an axial gauge with the polarization tensor
\beq
  {\cal P}^{\mu\nu}(k,n)
  = g^{\mu\nu} - {k^\mu n^\nu + n^\mu k^\nu \over n \cdot k},
\eeq
where the gauge vector $n$ satisfies $n^2=0$, and $n\,\cdot\, v=v^+/Q^+$.

Let us write the matrix element for light quark production
as $\bar{u}_\alpha(Q) {\cal M}_\alpha$, where $\alpha$ is a
Dirac index. The square of the amplitude for the full
$\eta_c$ production process can then be written as
${\cal M}^*_\beta {\cal T}_{\beta\alpha} {\cal M}_\alpha$.
In the limit of large $Q^+$,
\beq
  {\cal T}_{\beta\alpha} = T \Lslash{Q}_{\beta\alpha} + \dots
\eeq
where $T$ is a scalar function. The full cross section
then becomes a convolution of the light quark production
rate
\beq
  d\sigma = \left. \frac{1}{2E_{\rm CM}^2} \; {\rm dLips} \;
            {\cal M}^*_\beta \Lslash{Q}_{\beta\alpha} {\cal M}_\alpha 
            \right|_{Q^2 = 0}
\eeq
and a $q \to \eta_c$ fragmentation function which is given
by a phase space integral of $T$, as shown below.

\LevelTwo{The leading-order fragmentation function}

At lowest order, the $q \to \eta_c$ fragmentation function gets
contributions only from the Feynman diagram of Fig.\ \ref{momentumdef}
and another diagram where the two $c$-quark--gluon vertices have been
interchanged. The amplitudes corresponding to the two diagrams
are equal.

As shown in Fig.\ \ref{momentumdef}, we denote the momentum of
the virtual gluon by $k$ and define $w=k^2$ and $y=n \cdot k$.

Let us first consider the square of the $q^* \to q \eta_c$ amplitude
as it appears in the cross section of the full process. We find
\beqa
  {\cal T}
  & = & \frac{128\pi^2 \as^3 C_F \left| \RS \right|^2}{m_c N_c}
        \, {\Lslash{Q} \over s}
        \, \gamma^\beta \, (\Lslash{Q} - \Sslash{k}) \, \gamma^\alpha
        \, {\Lslash{Q} \over s}
        \nonumber \\
  &   & \hspace{.5 em} \times
        {{\cal P}_{\alpha\alpha'}(k,n) \over k^2}
        \; {p_\rho k_\xi \epsilon^{\rho\xi\alpha'\mu} \over k^2-4m_c^2}
        \; \left[-{\cal P}_{\mu\nu}(k-p,n)\right]
        \; {p_{\rho'} k_{\xi'} \epsilon^{\rho'\xi'\beta'\nu} \over k^2-4m_c^2}
        \; {{\cal P}_{\beta\beta'}(k,n) \over k^2}
        \nonumber \\
  & = & T_Q \Lslash{Q} + T_p \Lslash{p} + T_k \Lslash{k} + T_n \Lslash{n}
        \nonumber \\
  & = & (T_Q + zT_p + yT_k) \, \frac{1}{2} Q^+ \gamma^- + \dots
        \nonumber \\
  & = & (T_Q + zT_p + yT_k) \, \Lslash{Q} + \dots
  \label{quarkLineTree}
\eeqa
The tensor $p_\alpha k_\beta \epsilon^{\alpha\beta\mu\nu}$
is due to the $c\bar{c}$ spin projection \cite{Kuhn}.
The dots in the last two expressions stand for terms of relative order $1/Q^+$.
We made use of the fact that the coefficients $T_Q, T_p, T_k, T_n$
depend only on scalar products
of the four-momenta and are therefore independent of $Q^+$.
Explicitly,
\beqa
  T & = & T_Q + zT_p + yT_k \nonumber \\
    & = & \frac{128\pi^2 \as^3 C_F \left| \RS \right|^2}{m_c N_c}
          \nonumber \\
    &   & \times {1 \over 2 s^2 w^2 (w-4m_c^2)^2} \Biggl[
          -32\,{m_c^4}\,s - 16\,{m_c^4}\,w - 4\,\psq\,s\,w
          - 2\,s\,{w^2} - {w^3}
          \nonumber \\
    &   & \mbox{} + {{2\,s\,\left( 16\,{m_c^4} + 2\,\psq\,w + {w^2}
          \right)}\over y}
          + s\,\left( 16\,{m_c^4} + {w^2} \right) \,y
          - {{2\,{{\left( 4\,{m_c^2} + \psq \right) }^2}\,w}
          \over {{z^2}}}
          \nonumber \\
    &   & \mbox{} + {{2\,\left( 4\,{m_c^2} + \psq \right)
          \,w\,\left( 4\,{m_c^2} +  w \right) } \over z}
          - {{2\,s\,w\,\left( 4\,{m_c^2} + w \right) \,z} \over y}
          - 2\,{s^2}\,w\,{z^2}
          \nonumber \\
    &   & \mbox{} - {{2\,{s^2}\,w\,{z^2}} \over {y^2}}
          + {{2\,s\,w\,\left( 2\,s + w \right) \,{z^2}} \over y}
          \Biggr] .
\eeqa

The phase space measure for the full process can be written as
the product of three factors:
the phase space measure for light quark production,
the phase space measure for the decay of the virtual light quark,
and $ds/(2\pi)$. We write the two latter factors as
\beqa
  \skipfields
  \int {d s   \over (2 \pi)}
       {d^4 k \over (2 \pi)^4}
       {d^4 p \over (2 \pi)^4}
       2\pi\delta^+((Q-k)^2)
       2\pi\delta^+(p^2-4m_c^2)
       2\pi\delta^+((k-p)^2)
       \theta(\mu-s)
  \nonumber \\ \skipfields
  = \int_{\, 0}^{\, 1} dz
    \left[
    {1 \over 256 \pi^4}
    \int_{4m_c^2/z}^{\, \mu^2} \!\!\!\! ds
    \int_{\, z}^{\, 1} {dy \over y}
    \int_{4m_c^2 y/z}^{sy} \!\!\!\! dw
    \int_{-\sqrt{\rho}}^{\sqrt{\rho}}
    {dt \over \, \pi \sqrt{\rho-t^2}}
    \int {d \phi   \over (2 \pi)}
    \right]
  \label{PhaseSpaceMeasure}
\eeqa
where $\phi$ is the azimuthal angle of $\vec{p}_\bot$, and
\beqa
  \rho & = & {4z^2 \over y^4}(1-y)(y-z)(s\,y-w)
             \left( w \, z -4m_c^2 y \right) , \\
  t    & = & \psq + \left[
             {w\, z \over y^2}(2z - y - y\, z)
             + {z \over y}(s\, z-4m_c^2) + 4m_c^2 + s\, z^2 \right]  .
\eeqa
The integral over $z$ gives the
convolution in the production cross section, and the leading-order
light quark fragmentation function is
\beqa
  D_{{\!\!}_{q\to \eta_c}}(z,\mu)
  & = & 
% {\alpha_s^3 C_F \left| \RS \right|^2 \over 2\pi^2 m_c N_c}
  \frac{1}{256\pi^4}
        \int_{4m_c^2/z}^{\, \mu^2} \!\!\!\! ds
        \int_{\, z}^{\, 1} {dy \over y}
        \int_{4m_c^2 y/z}^{sy} \!\!\!\! dw
        \int_{-\sqrt{\rho}}^{\sqrt{\rho}}
        {T \, dt \over \, \pi \sqrt{\rho-t^2}}
        \nonumber \\
  & = & f(z) \ln \left( {\mu^2 \over 4m_c^2 } \right) + g(z) +
        {\cal O} \left( {4m_c^2 \over \mu^2} \right) +
        {\cal O}\!\left( \alpha_s^4 \right).
  \label{LOperturbativeD}
\eeqa
Analytical expressions for the functions $f(z)$ and $g(z)$ are given in
eqs.\ (\ref{fexpr}) and (\ref{gexpr}), respectively.

\LevelTwo{The loop contribution}

The trigger bias enhanced NLO contribution to the $q \to \eta_c$
fragmentation function comes from the Feynman diagram
in Fig.\ \ref{loopdef}a, and another diagram where the
two c-quark---gluon vertices have been interchanged.
The amplitudes corresponding to these two diagrams are equal.

The four-momenta are defined in Fig.\ \ref{loopdef}a.
The loop momentum is denoted by $k$, and $y=n \cdot k$.

We first consider the structure of the box loop integral
\beq
  \Gamma = \int {d^4 k \over (2 \pi)^4} \;
           {\gamma^\beta (\Lslash{Q} - \Sslash{k})
           \gamma^\alpha \over (Q-k)^2}
           \times {{\cal P}_{\alpha\mu}(k,n) \over k^2}
           \times {{\cal P}_{\beta\nu}(p-k,n) \over (p-k)^2}
           \times {p_\alpha k_\beta \epsilon^{\alpha\beta\mu\nu}
           \over (k^2-p\cdot k)} .
  \label{fullIntegral}
\eeq
The four factors in the integrand of eq.\ (\ref{fullIntegral})
are easily identified with the four sides of the box loop of
Fig.\ \ref{loopdef}.
Making a Dirac decomposition of the integrand we find
\beq
  \Gamma
  = \Lslash{A} + i \Lslash{B} \gamma^5
  = \int {d^4 k \over (2 \pi)^4} (\Sslash{a} + i \Sslash{b} \gamma^5) ,
\eeq
where
\beqa
  den \times a^\mu
  & = & \left[ -y\, (k\!\cdot\! p) - z\,
        (k\!\cdot\! Q)  + 2m_c^2\, y + {s\, y\over 2} + k^2z\right]
        k_\alpha n_\beta p_\gamma \epsilon^{\alpha\beta\gamma\mu}
        \nonumber \\
  &   & \mbox{} - z\, (k\!\cdot\! e)\, k^\mu +
        y\, (k\!\cdot\! e)\, p^\mu , \\
  den \times b^\mu
  & = & (1\! -\! y) \left[ 4m_c^2 y + (z - 2 y) (k\!\cdot\! p) \right] k^\mu
        \nonumber \\
  &   & \mbox{} - (1\! -\! y)
        \left[ y (k\!\cdot\! p) -k^2(2 y\! -\! z) \right] p^\mu
        \nonumber \\
  &   & \mbox{} + \Biggl[
        (4m_c^2 +s){ y\over 2} (k\!\cdot\! p)
        - y (k\!\cdot\! p)^2
        - 4m_c^2 y (k\!\cdot\! Q)
        \nonumber \\
  &   & \mbox{} + (2 y\! -\! z) (k\!\cdot\! p)(k\!\cdot\! Q)
        + k^2 \left( {s\, z\over 2} + 2m_c^2\, z - s\, y \right)
        \Biggr] \, n^\mu ,
\eeqa
and the denominators from the propagators and the gluon polarization
tensors are included in
\beq
  den = y (z-y) (Q-k)^2 k^2 (p-k)^2 \left[ k^2-(p\!\cdot\! k) \right] .
\eeq

Any four-vector $X^\mu$ can be written as a linear combination 
$X^\mu = X_n n^\mu + X_Q Q^\mu + X_p p^\mu + X_e e^\mu$ of
$n$, $Q$, $p$ and
$e^\mu = n_\alpha Q_\beta p_\gamma \epsilon^{\alpha\beta\gamma\mu}$.
It is easily seen that $a_p(k)$, $a_Q(k)$, $a_n(k)$, and $b_e(k)$
are all antisymmetric when $k$ is mirrored in the hyper 
plane spanned by $n, p$ and $Q$, \ie\ when $k_e \to -k_e$.
Therefore they do not contribute to the integral, and
\beqa
  A^\mu & = & A_e e^\mu = \int {d^4 k \over (2 \pi)^4} a_e(k) e^\mu , \\
  B^\mu & = & B_n n^\mu + B_Q Q^\mu + B_p p^\mu
              \nonumber \\
        & = & \int {d^4 k \over (2 \pi)^4}
              \left[ b_n(k) n^\mu + b_Q(k) Q^\mu + b_p(k) p^\mu \right] .
\eeqa
Analogously with eq.\ (\ref{quarkLineTree}), we now find
\beqa
  {\cal T} & = & \frac{\pi\as^4 \left| \RS \right|^2 C_F^2}{8N_c m_c^5} \;
                 {\Lslash{Q}\over s}
                 (\Lslash{A}^* - i\Lslash{B}^*\gamma^5)
                 (\Lslash{Q}-\Sslash{p})
                 (\Lslash{A} + i\Lslash{B}\gamma^5)
                 {\Lslash{Q}\over s}
                 \nonumber \\
           & = & K_Q \Lslash{Q} + K_p \Sslash{p} + K_n \Sslash{n}
                 + i K_{\Sslash{e}\gamma^5} \Sslash{e} \! \gamma^5
                 \nonumber \\
           & = & (K_Q + zK_p) \Lslash{Q} + \dots
                 \nonumber \\
           & = & \frac{\pi\as^4 \left| \RS \right|^2 C_F^2}{8N_c m_c^5} \;
                 {1-z \over s^2} |C|^2 \Lslash{Q} + \dots ,
\eeqa
where
\beqa
  C & = & (s\, z - 4m_c^2)A_e - (2B_n + s\, B_Q + s\, B_p)
          \nonumber \\
    & = & \int {d^4 k \over (2 \pi)^4} \Bigl[
          (s\, z - 4m_c^2)a_e - (2b_n + s\, b_Q + s\, b_p) \Bigr] .
          \label{C}
\eeqa
The phase space measure for the decay of the virtual light
quark times $ds/(2\pi)$ is in this case
\beq
  \frac{1}{32\pi^3}
  \int_0^1 dz
  \int_{4m_c^2/z}^{\mu^2} ds
  \int_0^{2\pi} d\phi .
\eeq
As $C$ is independent of the azimuthal angle $\phi$ of the decay,
we obtain the following 'box' contribution
to the light quark fragmentation function:
\beq
  D_{q\to \eta_c}^{\rm (box)}(z,\mu) = 
  \frac{\as^4 \left| \RS \right|^2 C_F^2}{128\pi N_c m_c^5}
  \int_{4m_c^2/z}^{\mu^2} ds {(1-z) \over s^2} |C|^2.
\eeq

We have only calculated the imaginary part of the box amplitude,
which is due to the sum
of the two Cutkosky cuts $I$ and $I\!I$ of Fig.\ \ref{loopdef}b.
This gives a lower limit of the full
loop contribution. The imaginary part is obtained by replacing, respectively,
\beqa
  \int {d^4 k \over (2 \pi)^4} & \stackrel{I}{\to} &
  {1 \over 2} \int {d^4 k \over (2 \pi)^4} \, k^2 \, (p-k)^2
  \, 2\pi \delta^+(k^2) \, 2\pi \delta^+((p-k)^2)
  \nonumber \\
  \int {d^4 k \over (2 \pi)^4} & \stackrel{I\!I}{\to} &
  {1 \over 2} \int {d^4 k \over (2 \pi)^4} \, k^2 \, (Q-k)^2
  \, 2\pi \delta^+(k^2) \, 2\pi \delta^+((Q-k)^2)
\eeqa
in eq. (\ref{C}). We find
\beqa
  {\rm Im}(C) & = & C^I + C^{I\!I}
                    = {4m_c^2 s \over s - 4m_c^2}
                    \left[ 1- {2s-4m_c^2 \over s-4m_c^2}
                    \ln\left(s \over 4m_c^2 \right) \right] , \\
  C^I         & = & {s\,z \over 1-z}
                    \ln \left( (s - 4m_c^2)z \over s\,z - 4m_c^2 \right) .
\eeqa
Performing the $s$ integration we get the result
given in eqs.\ (\ref{HOres},\ref{hexpr}).

%\pagebreak

\end{document}